\documentclass[twocolumn,notitlepage,prd,nofootinbib]{revtex4-1}
\usepackage{soul,xcolor,graphicx,amssymb,bm,hyperref}

\newcommand{\psr}{PSR~J0337+1715}
\newcommand{\rmd}{{\rm d}}

\begin{document}

\title{Testing the strong equivalence principle with the triple pulsar
  \psr{}}
\author{Lijing Shao}
\email{lshao@aei.mpg.de}
\affiliation{Max Planck Institute for Gravitational Physics (Albert Einstein
Institute), Am M\"uhlenberg 1, D-14476 Potsdam-Golm, Germany}
\date{\today}
\begin{abstract}
  Three conceptually different masses appear in equations of motion for objects
  under gravity, namely, the inertial mass, $m_{\cal I}$, the passive
  gravitational mass, $m_{\cal P}$, and the active gravitational mass, $m_{\cal
  A}$.  It is assumed that, for any objects, $m_{\cal I} = m_{\cal P} = m_{\cal
  A}$ in the Newtonian gravity, and $m_{\cal I} = m_{\cal P}$ in the Einsteinian
  gravity, oblivious to objects' sophisticated internal structure. Empirical
  examination of the equivalence probes deep into gravity theories. We study the
  possibility of carrying out new tests based on pulsar timing of the stellar
  triple system, \psr{}.  Various machine-precision three-body simulations are
  performed, from which, the equivalence-violating parameters are extracted with
  Markov chain Monte Carlo sampling that takes full correlations into account.
  We show that the difference in masses could be probed to $3\times10^{-8}$,
  improving the current constraints from lunar laser ranging on the
  post-Newtonian parameters that govern violations of $m_{\cal P}=m_{\cal I}$
  and $m_{\cal A}=m_{\cal P}$ by thousands and millions, respectively.  The test
  of $m_{\cal P}=m_{\cal A}$ would represent the first test of Newton's third
  law with compact objects.
\end{abstract}
\maketitle

\section{Introduction}

Mass is an important concept whose notion has evolved dramatically during
several important paradigm shifts in theoretical physics, from its original
meaning of {\it amount}, to {\it inertia} in Newtonian mechanics, to {\it
energy} in special relativity with the famous $E=mc^2$~\cite{ein05}.  Mass was
further developed by Einstein and Schwarzschild into an intimate relation with
the geometry of spacetime in general relativity (GR)~\cite{ein16,sch16a}.  In
quantum world, mass pertains to an object's de Broglie relation and Compton
wavelength in the nonrelativistic theory~\cite{bro23}.  In relativistic field
theories, the origin of mass results from spontaneous symmetry breaking with the
Higgs field seeking a minimum point of potential~\cite{hig14,eng14}, which was
verified at the LHC~\cite{aaa+12,cks+12}.  From a group-theoretic viewpoint,
mass is a Casimir invariant of the Poincar{\'e} group, hence labels the
irreducible representations~\cite{wig39}. 

We here study the concept of mass with the classical gravitational interaction.
In a theoretically independent way, there are three masses defined by
measurement~\cite{bon57}: i) the inertial mass, $m_{\cal I}$, enters Newton's
second law, $\bm{F}=m_{\cal I} \bm{a}$; ii) the passive gravitational mass,
$m_{\cal P}$, is the mass on which gravity acts, defined by $\bm{F} = - m_{\cal
P} \bm{\nabla} U$; iii) the active gravitational mass, $m_{\cal A}$, is the mass
that sources gravity, through the (integrated) Poisson's equation,
$\oint_{\partial V} \bm{g} \cdot \rmd \bm{A} = -4\pi G m_{\cal A}$.  In the
Newtonian gravity, these conceptually different masses are assumed to be equal,
namely $m_{\cal I} = m_{\cal P} = m_{\cal A}$. In GR, the geometric foundation
is built upon the equality of $m_{\cal I}$ and $m_{\cal P}$ (dubbed the
equivalence principle~\cite{will93}). The equality of $m_{\cal A}$ with the
other two is of debate in GR~\cite{bon92,rc92}.  While Bonnor found that,
assuming $m_{\cal I}=m_{\cal P}$, $m_{\cal A}$ deviates by a few times from
$m_{\cal P}$ for a static sphere of uniform density under strong
gravity~\cite{bon92}, Rosen and Cooperstock showed that there is only one mass
for an isolated body when the gravitational energy is taken into
account~\cite{rc92}.

The importance of experimental examination of equivalence of masses was realized
early in Newton's era~\cite{new87}. High precision tests of the weak equivalence
principle (i.e., $m_{\cal I} = m_{\cal P}$ for non-self-gravitating bodies)
include pendulum experiments of Newton, Bessel, Potter, and torsion-balancing
experiments of E\"otv\"os, Dicke, Braginsky, Adelberger, et al.~\cite{will14}.
Recent developments are putting the test into space with missions like
MICROSCOPE~\cite{tmlr12}, Galileo-Galilei~\cite{nsp+12}, and STEP~\cite{oewm12}.
In addition, lunar laser ranging (LLR)~\cite{bcd+73,wtb12} and pulsar
timing~\cite{ds91,sfl+05,gsf+11,fkw12,zsd+15} probed the equivalence principle
with self-gravitating bodies and limited the Nordtvedt parameter~\cite{nor68},
$\eta_{\rm N}$, to be less than $3\times10^{-4}$ and $3\times10^{-2}$
respectively.  In a vivid contrast, tests of the equality $m_{\cal P}= m_{\cal
A}$ are fewer. We only noticed two experiments,\footnote{In addition, Nordtvedt
  had a proposal to test $m_{\cal A}=m_{\cal P}$ by utilizing the Earth's
  south-north asymmetric distribution of ocean water~\cite{nor01}; but no
subsequent analysis is published.} one performed by Kreuzer using a Cavendish
balance that limited the difference in $m_{\cal P}/m_{\cal A}$ between fluorine
and bromine to  $\lesssim 5\times10^{-5}$~\cite{kre68}, and the other performed
by Bartlett and van Buren with LLR that limited the difference between iron and
aluminum to $\lesssim 4\times10^{-12}$~\cite{bb86}.

Here we propose new tests of equivalence of masses with the remarkable stellar
triple system, \psr{}~\cite{rsa+14}.  Various machine-precision three-body
simulations are performed closely following observational characteristics.
Possible violations in the equivalence of masses are {\it injected} directly via
equations of motion~\cite{lam11}, and recovered with a dedicated Markov chain
Monte Carlo (MCMC) sampler taking full correlations into account. Our results
suggest that the triple system has sensitivity $\sim 3\times10^{-8}$ to probe
the difference in masses. It could improve the current post-Newtonian limits by
thousands for $m_{\cal I} = m_{\cal P}$ and millions for $m_{\cal A} = m_{\cal
P}$, and would represent the first test of Newton's third law with compact
objects.

\section{The triple system}

\begin{figure}
  \centering
  \includegraphics[width=8cm]{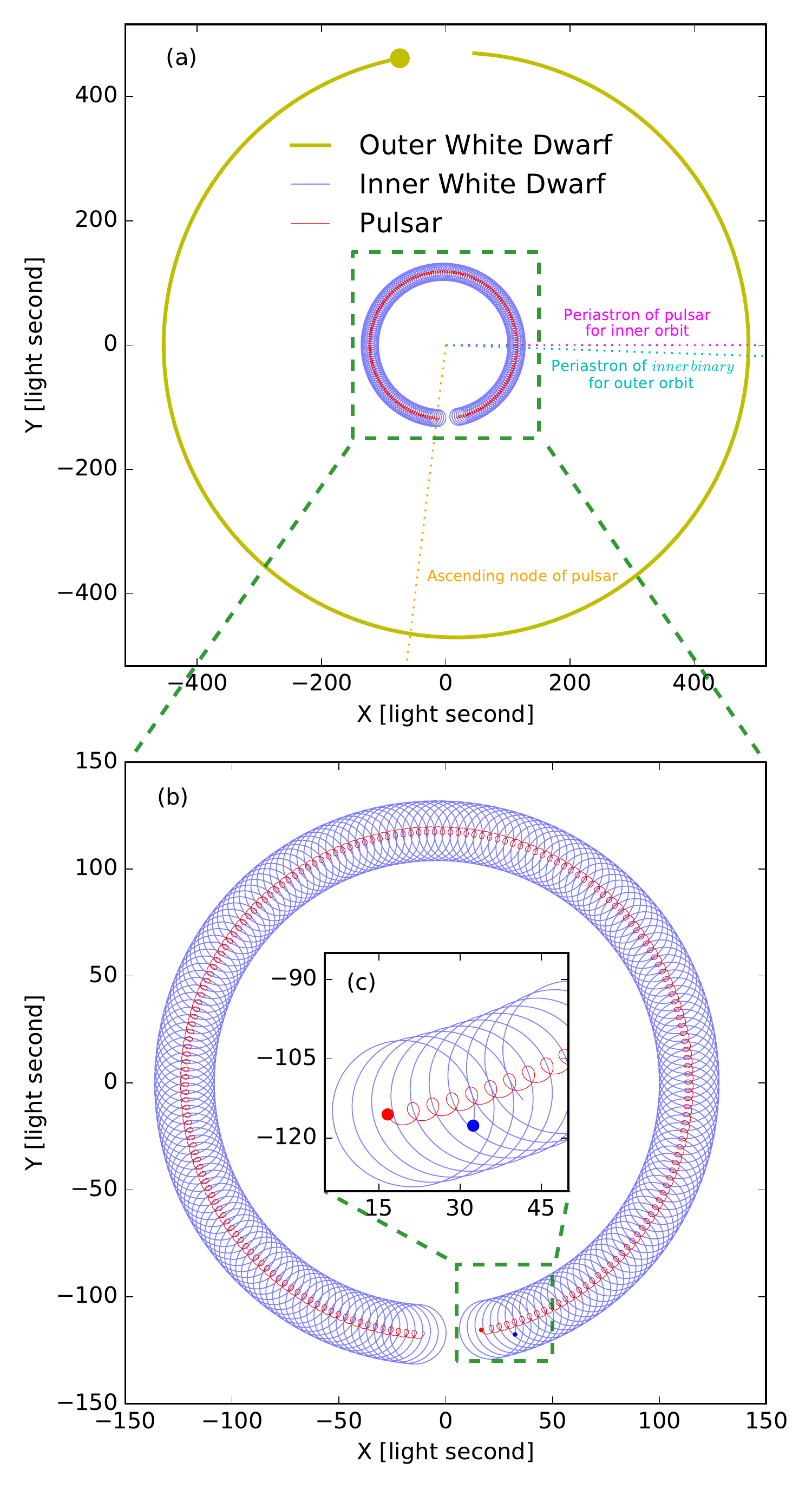}
  \caption{Illustration of the triple system, projected on the orbital plane of
    the {\it inner binary}. In (a) dotted lines mark directions of the
    periastron of the pulsar for the inner orbit, the periastron of {\it inner
    binary} for the outer orbit, and the ascending node of the pulsar; in Figure
    2 of Ref.~\cite{rsa+14}, these directions are indicated for WDs. (b) and (c)
    are magnified views of the regions enclosed by the green dashed boxes. These
    trajectories start on MJD 55920.0 (December 25, 2011), and end on MJD
  56233.9 (November 2, 2012). The starting locations are indicated by dots.}
  \label{fig:coplanar:orbit}
\end{figure}

\psr{} is a triple system consisting of a neutron star (NS) with mass
$1.44\,M_\odot$ and two white dwarfs (WDs) with masses $0.20\,M_\odot$ and
$0.41\,M_\odot$~\cite{rsa+14,th14}. The NS and the lighter WD are
gravitationally bound as an {\it inner binary} with $P_{\rm b,I} = 1.63$\,d that
are, as a whole, hierarchically bound to the outer WD with $P_{\rm b,O} =
327$\,d. Two orbits are very circular with $e_{\rm I}=6.9\times10^{-4}$ for the
{\it inner binary}, and $e_{\rm O}=3.5\times10^{-2}$ for the outer orbit. Two
orbital planes are remarkably coplanar with an inclination $\lesssim
0.01^\circ$~\cite{rsa+14}.  

An illustration of orbits is given in Figure~\ref{fig:coplanar:orbit}.  It was
simulated with the parameters reported in Ref.~\cite{rsa+14}. Initial conditions
are worked out for MJD 55920.0 which is the reference epoch for all parameters.
The three-body evolution under Newtonian gravity is performed with the {\sc
ias15} integrator in {\sc
rebound}\footnote{\url{https://github.com/hannorein/rebound}}~\cite{rl12}.  The
{\sc ias15} integrator is a 15th-order integrator based on the Gau\ss-Radau
quadrature.  It uses adaptive time stepping, and keeps systematic errors well
below machine precision over $10^9$ orbits~\cite{rs15}. The precision is very
important for three-body dynamics, because the pulsar timing experiments
spanning $\sim1.4$\,yr ($\sim4\times10^7$\,s) have achieved a weighted RMS
residual, $\sigma_{\rm TOA} = 1.34\,\mu$s~\cite{rsa+14}.  Our numerical
integration has to be more accurate than that in order to study tiny effects in
the orbital dynamics.

\section{Pulsar timing and parameter estimation}

We evolve the triple system in 3D for a longer time than the observation in
Ref.~\cite{rsa+14}, and then cut data keeping the part which corresponds to the
real data span (MJD 55930.9---56436.5). A spin-down model for the pulsar, $f(t)
= f_0 + \dot f t$, is constructed with a spin frequency,
$f_0=365.953363096\,$Hz, and its first time derivative, $\dot
f=-2.3658\times10^{-15}\,{\rm Hz}\,{\rm s}^{-1}$.  By projecting the pulsar's
trajectory along its line of sight to the Earth, we obtain the geometric delay
of pulse signals (i.e., the R\"omer delay).  Together with the spin-down model,
simulated times of arrival (TOAs), $N(t)$, with $N$ the counting number of
pulses and $t$ the coordinate time, are recorded. Relativistic effects (e.g.,
the periastron advance, the gravitational damping, the Shapiro time delay) are
not observable in practice yet~\cite{rsa+14}, therefore not included. The only
exception is the transverse Doppler effect due to the cross term of velocities
for inner and outer orbits~\cite{rsa+14}. It is approximated as $R(t) \simeq
\int \frac{1}{c^2} \bm{v}_{\rm O} \cdot \bm{v}_{\rm I} \rmd t = \frac{1}{c^2}
\bm{x}_{\rm I}(t) \cdot \bm{v}_{\rm O}(t) - \int \frac{1}{c^2} \bm{x}_{\rm I}
\cdot \rmd \bm{v}_{\rm O} \simeq \frac{1}{c^2} \bm{x}_{\rm I}(t) \cdot
\bm{v}_{\rm O}(t)$, where constants and the integral term, which is smaller by a
factor $\sim P_{\rm b,I}/P_{\rm b,O}$ on the timescale of the inner orbit, are
dropped~\cite{bt76}. $R(t)$ has an amplitude $\sim50\,\mu$s, consistent with the
real data~\cite{rsa+14}.  26280 TOAs are sampled from our simulation either
uniformly in time ({\it uniform sampling} hereafter) or with fake observing
blocks once a week with TOAs being separated by 10 seconds within block ({\it
step sampling}).  A Gaussian noise with a variance $\sigma_{\rm TOA} =
1.34\,\mu{\rm s}$ is added homogeneously to TOAs to mimic the observation
uncertainty~\cite{rsa+14}. Several noise realizations are simulated for each
sampling method.  

\begin{figure}
  \centering
  \includegraphics[width=9cm]{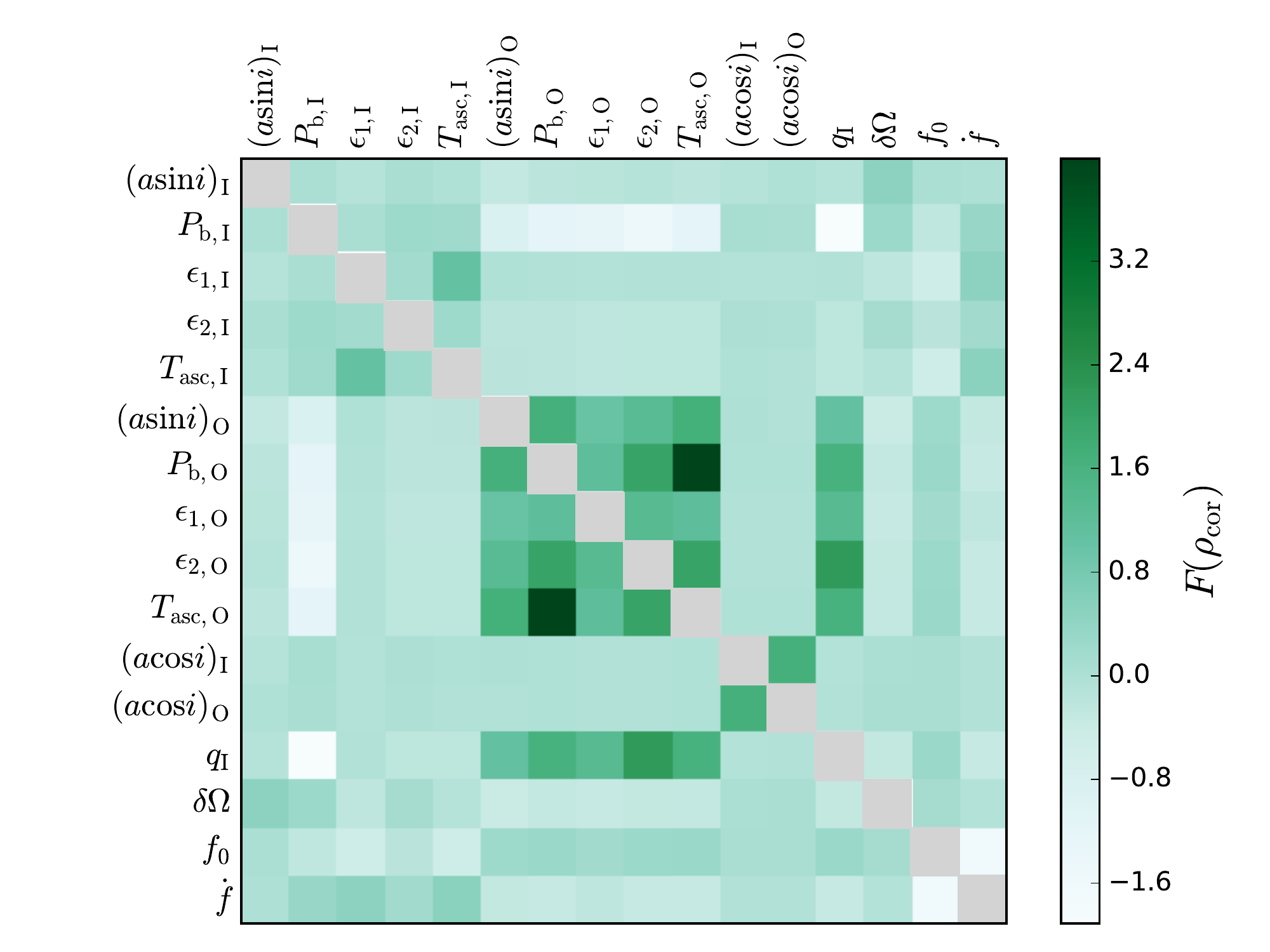}
  \caption{The correlations between 16 fitting parameters.  $F(\rho_{\rm cor})
  \equiv \log_{10}[(1+\rho_{\rm cor})/(1-\rho_{\rm cor})] - \rho_{\rm
  cor}\log_{10}2$ is a function defined such that it counts $9$'s in the limit
  of large correlations [e.g., $F(0.999)\simeq+3$, $F(-0.9) \simeq -1$, and
    $F(0) = 0$]; on diagonal, $F(\rho_{\rm cor})$ diverges.}
  \label{fig:corr:uniform}
\end{figure}

Following the method in Ref.~\cite{rsa+14}, we set up MCMC runs to estimate
parameters. To follow the fitting of real data as closely as possible, the same
set of parameters are used, which include 2 parameters for the pulsar's
spin-down, and 14 parameters for the size, the shape, the orientation, and the
initial condition of two orbits (details can be found in Ref.~\cite{rsa+14}).
The {\sc python} implementation of an affine-invariant MCMC ensemble
sampler~\cite{gw10,fhlg13}, {\sc emcee},\footnote{\url{http://dan.iel.fm/emcee}}
is used to explore the 16D parameter space.   In each step, we generate
noiseless template TOAs according to 16 parameters that are being sampled by the
kernel. They are compared with the TOAs generated before.  The runs proceed the
exploration of parameter space according to the difference between two sets of
TOAs, characterized by $\chi^2$ (for details of the Markov-chain implementation,
see Ref.~\cite{fhlg13}).

We accumulate $320000$ MCMC samples for each set of simulated TOAs, of which the
first half are abandoned as the {\sc burn-in} phase~\cite{bgjm11}. The
Gelman-Rubin statistic is used to verify the convergence of different
chains~\cite{gr92}. The 16D parameter space is marginalized to obtain the
uncertainty for each parameter.  It is remarkable that with the only input of
the orbital characteristics and a timing noise, we recover all observational
{\it uncertainties} for 14 orbital parameters~\cite{rsa+14} within a factor of
2, except the difference in the longitude of ascending nodes for two orbits,
whose uncertainty is off by a factor of 3. Uncertainties of the spin-down
parameters are however underestimated, by factors of $4000$ for $f_0$ and $10$
for $\dot f$, which could be caused by our simplified sampling method. It is
interesting to note that the uncertainties of $f_0$ and $\dot f$ are relatively
large for \psr{}, by factors of $10^4$--$10^5$, when compared with binary
pulsars of similar high-quality observations with a comparable span, the number
of TOAs, and the timing residual; see e.g., PSRs~J0737$-$3039A~\cite{ksm+06} and
J0348+0432~\cite{afw+13}.  The correlations between 16 parameters are plotted in
Figure~\ref{fig:corr:uniform} for simulated TOAs with {\it uniform sampling}.
The largest correlation comes from the time of ascending node and the orbital
period for the outer orbit which, we suspect, is related to the small number
($\sim1.5$) of orbital coverage, that makes the variables of the outer orbit
likely correlated (see the green $5\times5$ sub-block in
Figure~\ref{fig:corr:uniform}).  The correlation matrices for different noise
realizations are hardly distinguishable, and those for {\it step sampling} are
fully consistent with Figure~\ref{fig:corr:uniform}.

\section{Equivalence of masses}

\begin{figure}
  \centering
  \includegraphics[width=8cm]{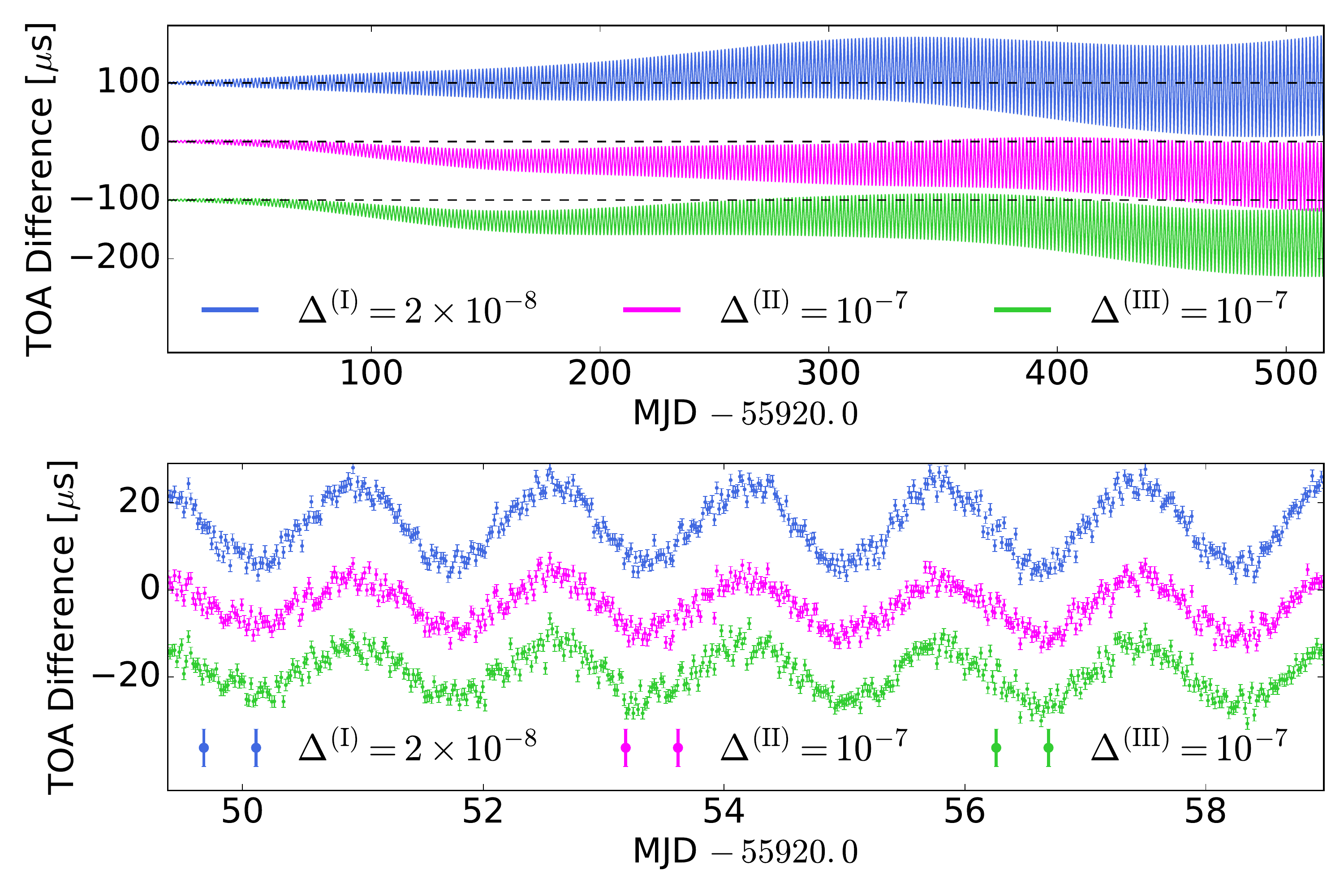}
  \caption{The difference in simulated TOAs introduced by hypothetical
  violations in the equivalence of masses. Lower panel is a magnified view of a
small region that contains 6 inner orbits. The blue and green series are offset
vertically for a better view.}
  \label{fig:diff:grav}
\end{figure}
 
The discovery of the triple pulsar has triggered some studies in tests of the
strong equivalence principle (i.e. $m_{\cal I} = m_{\cal P}$ for
self-gravitating bodies)~\cite{rsa+14,will14}. Preliminary results showed that
it probes the difference in $m_{\cal P}/m_{\cal I}$ between NSs and WDs at
$10^{-5}$--$10^{-8}$~\cite{hl14,ssa+15}. The Square Kilometre Array will improve
that further and limit the scalar-tensor gravity stringently~\cite{bbc+15}.  No
detailed analysis has been published yet. Here we perform such a study. In
addition to $m_{\cal I} = m_{\cal P}$, a new test is proposed to study the
possibility of $m_{\cal A} \neq m_{\cal P}$.  Because Newton's third law is
violated if $m_{\cal A} \neq m_{\cal P}$~\cite{bon57}, it is the first test of
the famous $actio = reactio$ formalism with strongly self-gravitating bodies.

The two WDs in the triple system are assumed to have a similar strength in
violating the equivalence of masses.\footnote{It is easy to relax this
assumption, but leading to an unnecessary redundancy with little theoretical
interests.} The equivalence-violating parameters are defined as, $\Delta^{\rm
(I)}\equiv\left({m_{\cal A}}/{m_{\cal I}}\right)_{\rm NS} - 1$, $\Delta^{\rm
(II)}\equiv\left({m_{\cal P}}/{m_{\cal I}}\right)_{\rm NS} - 1$, $\Delta^{\rm
(III)}\equiv\left({m_{\cal A}}/{m_{\cal I}}\right)_{\rm WD} - 1$, and
$\Delta^{\rm (IV)}\equiv\left({m_{\cal P}}/{m_{\cal I}}\right)_{\rm WD} - 1$.
Corresponding modifications to the gravitational interaction are added to the
{\sc ias15} integrator~\cite{rs15,rl12}, via
\begin{equation} 
m_{i,{\cal I}} \frac{\rmd^2 \bm{r}_i}{\rmd t^2}  \equiv
m_{i,{\cal I}} \bm{a}_i = \sum_{j\neq i} - \frac{G m_{i,{\cal P}} m_{j,{\cal
A}} }{r_{ij}^3} \bm{r}_{ij} \,, 
\label{eq:force} 
\end{equation}
with $\bm{r}_{ij} \equiv \bm{r}_i - \bm{r}_j$ and $r_{ij} \equiv |\bm{r}_{ij}|$.
Further analysis shows that one $\Delta$ can be set to vanish, which is related
to an unobservable rescaling.  We choose to set $\Delta^{\rm (IV)} = 0$.
Consequently, the remaining three $\Delta$'s should be interpreted as the
difference in the mass ratio relative to $\left({m_{\cal P}}/{m_{\cal
I}}\right)_{\rm WD}$.

Figure~\ref{fig:diff:grav} shows examples of the difference in simulated TOAs
with the equivalence violation, with respect to TOAs that are simulated with
Newtonian gravity.  With $\Delta$'s of $10^{-8}$--$10^{-7}$, the effects on TOAs
are already {\it much} larger than the achieved timing residual. However, the
correlation with orbital elements is strong. In order to assess the true
sensitivity of the triple pulsar, a simultaneous fitting of $\Delta$'s with
other parameters is {\it necessary}.

\begin{figure}
  \centering
  \includegraphics[width=9cm]{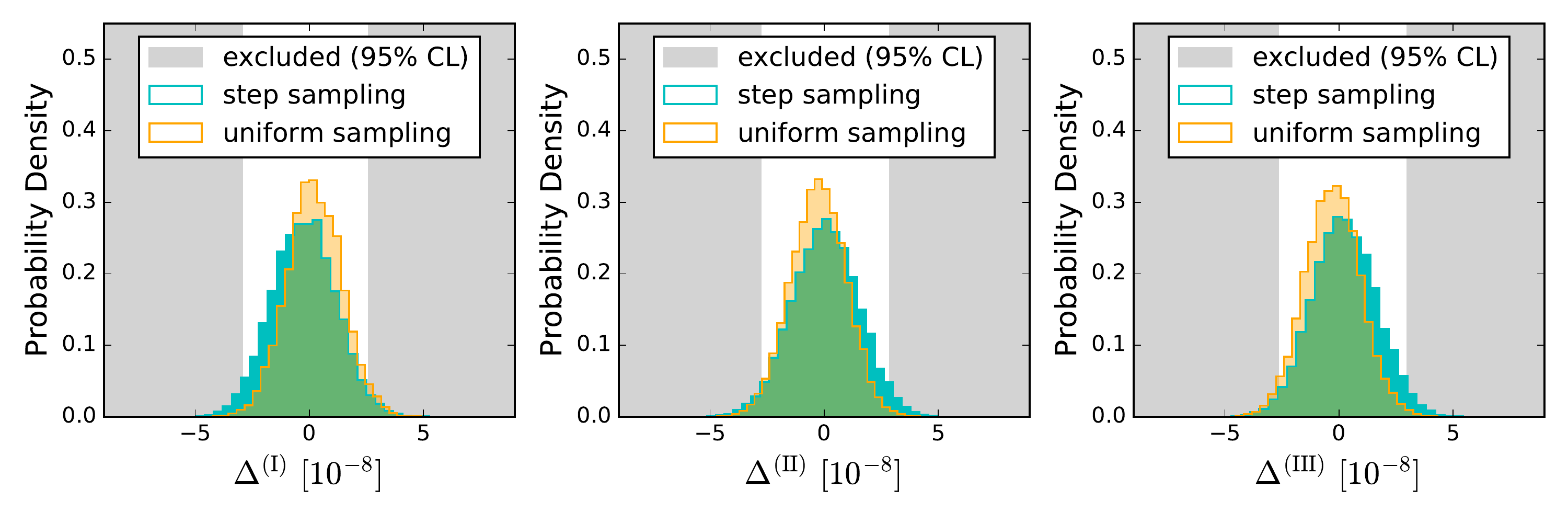}
  \caption{The {\it posteriori} PDFs for the equivalence-violating parameters
  from simulated TOAs with
  $\Delta^{\rm(I)}=\Delta^{\rm(II)}=\Delta^{\rm(III)}=0$.}
  \label{fig:hist:noinjection}
\end{figure}

We probe the sensitivity of \psr{} in constraining $\Delta$'s by adding a
nonzero $\Delta$ in the parameter-estimation process.  Fake TOAs are simulated
as before. Template TOAs are generated with the possibility of allowing a
nonvanishing $\Delta$.  Because of the strong mutual correlations (see
Figure~\ref{fig:diff:grav}), we are not able to estimate three $\Delta$'s at one
time.\footnote{Simultaneous fittings with three $\Delta$'s are tried, but the
  convergence is very bad after a long MCMC run.}  Instead, they are analyzed
  separately. 320000 MCMC samples are accumulated for each set of simulated TOAs
  for each $\Delta$.  After dropping the first half {\sc burn-in} runs and
  marginalizing over 16 parameters, we obtain the {\it posteriori} probability
  density functions (PDFs) for $\Delta$'s.  Different noise realizations give
  consistent results.  One example is shown in
  Figure~\ref{fig:hist:noinjection}.  The region that is excluded by both
  sampling methods is conservatively taken as the exclusion region. We conclude
  that, the data quality of \psr{} presented in Ref.~\cite{rsa+14} allows one to
  constrain $|\Delta|$'s to $\lesssim 3\times10^{-8}$. Because, as seen from
  Eq.~(\ref{eq:force}), all $\Delta$'s modify the trajectories in a similar way,
  it is not surprising that they are to be constrained with a similar precision.

\begin{figure}
  \centering
  \includegraphics[width=9cm]{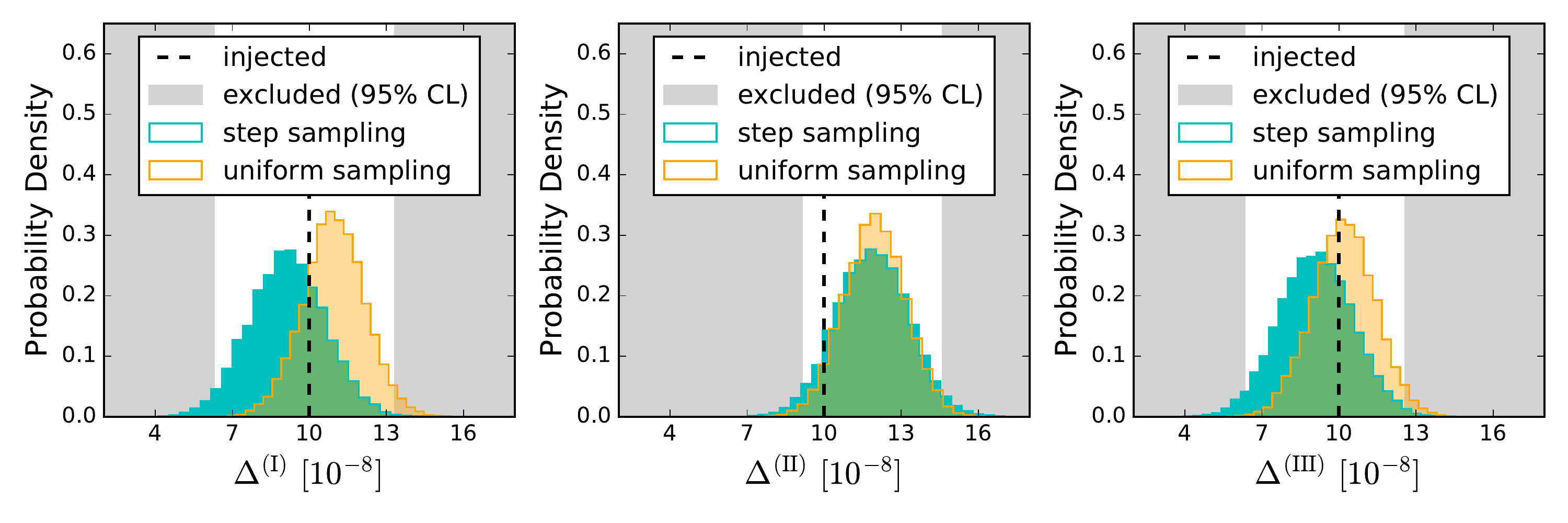}
  \caption{The recovery of $\Delta$'s from simulated TOAs with
    $\Delta^{\rm(I)}=10^{-7}$ (left), $\Delta^{\rm(II)}=10^{-7}$ (middle), and
    $\Delta^{\rm(III)}=10^{-7}$ (right).} \label{fig:hist:nonzero}
\end{figure}

In addition to {\it constrain} the equivalence violation in masses, the
capability of \psr{} to {\it detect} such violations, if they indeed exist, is
also investigated. We {\it inject} nonvanishing $\Delta$'s into our simulated
TOAs by modifying the orbital dynamics according to Eq.~(\ref{eq:force}). The
same parameter-estimation process by allowing one nonzero $\Delta$ is performed
with these TOAs. The resulting {\it posteriori} PDFs are shown in
Figure~\ref{fig:hist:nonzero}. As one can see, the equivalence violation can be
detected if it indeed exists.

\section{Discussions}

The equivalence of masses is vital to gravity theories. Already with the {\it
metric theories of gravity} that fulfill the Einstein's equivalence
principle~\cite{will93}, three conceptually different masses are
distinguishable.  For example, in the parametrized post-Newtonian (PPN)
formalism~\cite{will76a,will93,will14}, \small
\begin{equation}
  \frac{m_{\cal P}}{m_{\cal I}} = 1 - \left( 4\beta - \gamma - 3
  - \alpha_1 - \frac{2}{3} \zeta_1 -
  \frac{1}{3} \zeta_2 \right) \frac{E_{\cal G}}{m_{\cal I}c^2} \,,
  \label{eq:passive}
\end{equation}
\begin{equation}
  \frac{m_{\cal A}}{m_{\cal I}} = 1 - \left( 4\beta - \gamma - 3 - 2\zeta_2 -
  \frac{1}{3} \zeta_1 \right) \frac{E_{\cal G}}{m_{\cal I}c^2} \,,
  \label{eq:active}
\end{equation}
\normalsize where we have set PPN parameters $\alpha_2 = \alpha_3 = \xi = 0$,
due to their tight limits ($10^{-9}$ for $|\alpha_2|$,
$|\xi|$~\cite{sck+13,sw13}; $10^{-20}$ for $|\alpha_3|$~\cite{sfl+05}).  Using
$E_{\cal G} / m_{\cal I}c^2 \simeq 0.1 m_{\rm NS}/M_\odot$ for NSs~\cite{de92a},
one constrains $\left| 4\beta - \gamma - 3 - \alpha_1 - \frac{2}{3} \zeta_1 -
\frac{1}{3} \zeta_2 \right|$ and $\left| 4\beta - \gamma - 3 - 2\zeta_2 -
\frac{1}{3} \zeta_1 \right|$ to $\lesssim 2\times10^{-7}$, with the limits on
$\Delta^{\rm (I)}$ and $\Delta^{\rm (II)}$ from \psr{}.\footnote{For this
  particular analysis, $\Delta^{\rm(III)}$ can be assumed to vanish, due to the
  relatively weak gravity of WDs, $\left(E_{\cal G}/m_{\cal I}c^2\right)_{\rm
  WD} \simeq 10^{-4}$.} Without a fortuitous cancellation, $\beta$, $\gamma$,
  $\alpha_1$, $\zeta_1$, and $\zeta_2$, can be constrained to $\lesssim
  10^{-7}$, improving the current best bounds~\cite{will14} by $10^2$--$10^5$.
  Even allowing a fortuitous cancellation, one still improves their bounds, for
  example, {\it at least} by $\gtrsim10^3$ for $\zeta_1$. 

With the limit on $\Delta^{\rm(II)}$, the Nordtvedt parameter~\cite{nor68},
$\eta_{\rm N}$ ($= 4\beta-\gamma-3 -\alpha_1 - \frac{2}{3}\zeta_1 -
\frac{1}{3}\zeta_2$ in the PPN formalism~\cite{will93}), improves by
$\gtrsim10^3$ with respect to LLR~\cite{wtb12}.  This would be the first time
that compact objects provide a tighter limit on $\eta_{\rm N}$ than the Solar
system.  The test of $m_{\cal P} = m_{\cal A}$ would be the first test with
strongly self-gravitating bodies, which vastly extends the regime explored by
the previous tests in terms of objects' compactness~\cite{kre68,bb86}. The test
would surpass the best test~\cite{bb86} by $10^6$ within the post-Newtonian
analysis, and would be the first test of Newton's third law with strongly
self-gravitating bodies.

We stress that, although our simulated TOAs are able to reproduce major features
of the real observation~\cite{rsa+14}, they are simplified compared with the
complications in the real data, e.g., the heteroscedasticity in TOAs from
different telescopes,  the irregular jumps between observing sessions, the
remove of time-dependent interstellar dispersion, the correlation with parallax
and proper motion~\cite{kop96}. This study is intended to advocate the program
to analyze foundational principles on the equivalence of masses with the
remarkable triple system.  The analysis in this work is solely based on the
observation presented in Ref.~\cite{rsa+14}. In reality, more data have
accumulated since that publication.  We urge observers to test the equivalence
of masses with real timing data.

\begin{acknowledgments}
We thank Stanislav Babak, Alessandra Buonanno, Abraham Harte, Vivien Raymond,
and Norbert Wex for helpful discussions. The Markov chain Monte Carlo runs were
performed on the {\sc vulcan} cluster at the Albert Einstein Institute in
Potsdam-Golm.
\end{acknowledgments}

\end{document}